\documentclass[10pt,a4paper]{article}
\usepackage[dvips]{color}
\usepackage{epsfig}
\usepackage{amsmath}
\usepackage{graphicx}
\usepackage{authblk}
\usepackage{amssymb,amsmath}
\usepackage{amsmath}

\begin{document}
\title{\bf Interacting Quintessence Models of Dark Energy}
\author{ {M. Khurshudyan$^{a}$ \thanks{Email: khurshudyan@yandex.ru}}, \hspace{1mm} E. Chubaryan$^{a}$ \thanks{Email: echub@ysu.am}, \hspace{1mm} and B. Pourhassan $^{b, c}$\thanks{Email: b.pourhassan@umz.ac.ir}\\
$^{a}${\small {\em Department of Theoretical Physics, Yerevan State
University, 1 Alex Manookian, 0025, Yerevan, Armenia}}\\
$^{b}${\small {\em Department of Physics, Damghan
University, Damghan, Iran}}\\
$^{c}${\small {\em Physics Department, Istanbul Technical University, Istanbul, Turkey}}} \maketitle
\begin{abstract}
In this paper we consider two models of quintessence scalar fields with different potentials. Interaction with generalized cosmic Chaplygin gas is also investigated. Cosmological parameters are studied and graphical behavior is analyzed. We find that our model is agree with observational data specially $\Lambda$CDM model.
{\bf Pacs Number(s):} 95.35.+d, 95.85.-e, 98.80.-k
\end{abstract}

\section{\large{Introduction}}
Conception of an accelerated expansion, explanation of the phenomenon within dark energy, thought to have an inflation, which can solve the question of flatness, are all model based thoughts, seeming phenomena of the absolute observer. Over the years we are trying to solve problems of modern cosmology, just citing to the ideas that nature of the dark energy, nature of dark mater, nature of interaction between components are unknown. This fact makes possibility to huge number of speculations, with different class of outcomes. One of the outcomes is a possibility of the future singularities, which makes us really afraid. From the one hand we have singularities at the beginning of our Universe and from the other hand we have other singularities at the end. For future singularities we have theoretical framework to distinguish them from each other, and theoretically with some certainty to suppose possible origin of them. However, singularities of the beginning in the history are not distinguishable (so far we know). We account them as a one single concept.  Among different contradictory facts, it seems that cosmologists created to different groups believing to different facts, but careful investigation of the origin of the believes, we can find the same believe seated at the base with different formulations. Recall, that a set of observational data reveal that an expansion of our Universe is accelerated [1-3]. Then, the density of matter is very much less than critical density [4], the Universe is flat and the total energy density is very close to the critical. Explanation of accelerated expansion of our Universe in modern cosmology is based on an idea of dark energy. Among different viewpoints concerning to the nature of the dark component of the Universe, we would like to mention a scalar field models, one of them is Tachyonic field with its relativistic Lagrangian,
\begin{equation}\label{eq:tach lag}
L_{TF}=-V(\phi)\sqrt{1-\partial_{\mu}\phi\partial^{\nu}\phi},
\end{equation}
which captured a lot of attention.
The stress energy tensor,
\begin{equation}\label{eq:energy tensor}
T^{ij}=\frac{\partial L}{\partial (\partial_{i}\phi)}\partial^{k}\phi-g^{ik}L,
\end{equation}
gives the energy density and pressure as,
\begin{equation}\label{eq:tachyonic density}
\rho=\frac{V(\phi)}{\sqrt{1- \partial_{i}\phi \partial^{i}\phi}},
\end{equation}
and,
\begin{equation}\label{eq:tachyonic pressure}
P=-V(\phi)\sqrt{1- \partial_{i}\phi \partial^{i}\phi}
\end{equation}
A quintessence field (which is under our consideration in this work) is other model based on scalar field with standard kinetic term, which minimally coupled to gravity. In that case the action has a wrong sign kinetic term and the scalar field is called phantom. Combination of the quintessence and the phantom is known as the quintom model. Extension of kinetic term in Lagrangian yields to a more general frame work on field theoretic dark energy, which is called k-essense. A singular limit of k-essense is called Cuscuton model.
This model has an infinite propagating speed for linear perturbations, however causality is still valid. The most general form for a scalar field with second order equation of motion is the Galileon field which also could behaves as dark energy. See Refs. [5-11] for several models of dark energy.\\
An interesting model of dark energy is based on Chaplygin gas (CG) equation of state. The CG was not
consistent with observational data and extended to generalized Chaplygin gas
(GCG) [12-14], and indeed proposed unification of dark matter and dark
energy. It is also possible to study viscosity in GCG [15-20]. However, observational data ruled out such a proposal and,
GCG extends to the modified Chaplygin gas (MCG) [21]. Recently, viscous MCG is also suggested and investigated [22]. One of the last extensions of CG models is generalized cosmic Chaplygin gas (GCCG) which has been proposed recently [23].\\
In this work we will consider two models of quintessence scalar field interacting with generalized cosmic Chaplygin gas. For both models we assumed that potential of the field is given. For the first model,
\begin{equation}\label{eq:firstpot}
V(\phi)=\exp{( - \beta \phi)}.
\end{equation}
Second interacting model will be described by the following potential,
\begin{equation}\label{eq:secpot}
V(\phi)=\phi^{-2}+\rho_{1}\exp{( - \beta \phi)}.
\end{equation}
For a given quintessence scalar field with a potential $V(\phi)$ we can associate energy density and pressure of an ideal fluid as,
\begin{equation}\label{eq:endenq}
\rho_{Q}=\frac{1}{2}\dot{\phi}^{2}+V(\phi),
\end{equation}
and,
\begin{equation}\label{eq:endenq}
P_{Q}=\frac{1}{2}\dot{\phi}^{2}-V(\phi),
\end{equation}
with the following EoS parameter,
\begin{equation}\label{eq:EoSQ}
\omega_{Q}=\frac{\dot{\phi}^{2}-2V(\phi)}{\dot{\phi}^{2}+2V(\phi)}.
\end{equation}
Interaction between components is assumed to be
\begin{equation}\label{eq:int}
Q=(3Hbq+\frac{\dot{\phi}}{\phi}) \rho_{1},
\end{equation}
where $q$ is a deceleration parameter, $\rho_{1}$ is the energy density of generalized cosmic Chaplygin gas and EoS of it is written as,
\begin{equation}\label{eq:GCCG}
P_{1}=\gamma \rho_{1}-\frac{1}{\rho_{1}^{\alpha}}\left( A+ (\rho_{1}^{1+\alpha}-A)^{-\omega}\right),
\end{equation}
where $A$ assumed as a function of potential $V(\phi)$,
\begin{equation}\label{eq:A}
A(\phi)=\frac{V(\phi)}{1-\omega}-1.
\end{equation}
It is indeed an interesting idea because the coefficient $A$ already considered as a constant or a scale-factor dependent coefficient.
The paper organized as follow: in next section we will introduce the equations which governs our model. In section 3 we give numerical analysis and discuss about cosmological parameters in two different models of given potential. Finally we give conclusion.

\section{\large{The field equations and Models}}
Field equations that govern our model of consideration are,
\begin{equation}\label{eq:Einstein eq}
R^{\mu\nu}-\frac{1}{2}g^{\mu\nu}R^{\alpha}_{\alpha}=T^{\mu\nu}.
\end{equation}
By using the
following FRW metric for a flat Universe,
\begin{equation}\label{s2}
ds^2=-dt^2+a(t)^2\left(dr^{2}+r^{2}d\Omega^{2}\right),
\end{equation}
field equations can be reduced to the following Friedmann equations,
\begin{equation}\label{eq: Fridmman vlambda}
H^{2}=\frac{\dot{a}^{2}}{a^{2}}=\frac{\rho}{3},
\end{equation}
and,
\begin{equation}\label{eq:Freidmann2}
\dot{H}=-\frac{1}{2}(\rho+P),
\end{equation}
where $d\Omega^{2}=d\theta^{2}+\sin^{2}\theta d\phi^{2}$, and $a(t)$
represents the scale factor. The $\theta$ and $\phi$ parameters are
the usual azimuthal and polar angles of spherical coordinates, with
$0\leq\theta\leq\pi$ and $0\leq\phi<2\pi$. The coordinates ($t, r,
\theta, \phi$) are called co-moving coordinates.\\
Energy conservation $T^{;j}_{ij}=0$ reads as,
\begin{equation}\label{eq:Bianchi eq}
\dot{\rho}+3H(\rho+P)=0.
\end{equation}
To introduce an interaction between components (\ref{eq:Bianchi eq}) we should mathematically split it into two following equations
\begin{equation}\label{eq:inteqm}
\dot{\rho}_{1}+3H(\rho_{1}+P_{1})=Q,
\end{equation}
and,
\begin{equation}\label{eq:inteqG}
\dot{\rho}_{Q}+3H(\rho_{Q}+P_{Q})=-Q.
\end{equation}
Cosmological parameters of our interest are EoS parameters of each fluid components $\omega_{i}=P_{i}/\rho_{i}$, EoS parameter of composed fluid,
\begin{equation}\label{20}
\omega_{tot}=\frac{P_{Q}+P_{\Lambda} }{\rho_{Q}+\rho_{\Lambda}},
\end{equation}
deceleration parameter $q$, which can be written as,
\begin{equation}\label{eq:accchange}
q=\frac{1}{2}(1+3\frac{P}{\rho} ),
\end{equation}
where $P=P_{Q}+P_{1}$ and $\rho=\rho_{Q}+\rho_{1}$, which is nested at the base of modern theoretical cosmology. Such assumption gives us possibility to account only non minimal couplings, simplifies calculations. Consideration of an interaction between components in the form of $Q$ is a speculative job. The number of the forms of $Q$ therefore could accept infinite number, even observations could not help us to fix them with a hope to decrease number. Moreover, new observations support to the possibility to increase different possibilities to increase the number of different possible forms of $Q$.  One of the examples is considered in our work, where we include deceleration parameter to provide sign changeability to one of the terms.
\section{\large{Numerical Results and Cosmological parameters}}
\subsection{\large{Model 1}}
First model is based on the assumption that potential of the scalar field $V(\phi)$ has an exponential connection with field given by the equation (5). We give numerical analysis to obtain some cosmological parameters which illustrated in the Figs. 1-6. Fig. 1 shows that Hubble expansion parameter is decreasing function of time which is expected. We can see that increasing $\gamma$ decreases value of Hubble expansion parameter.\\
\begin{figure}[h!]
 \begin{center}$
 \begin{array}{cccc}
\includegraphics[width=70 mm]{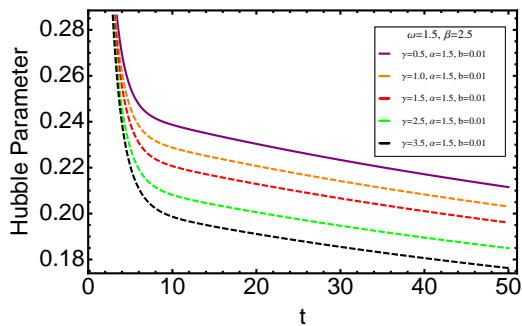}
 \end{array}$
 \end{center}
\caption{Behavior of Hubble parameter $H$ against $t$ for model 1.}
 \label{fig:1}
\end{figure}

\begin{figure}[h!]
 \begin{center}$
 \begin{array}{cccc}
\includegraphics[width=70 mm]{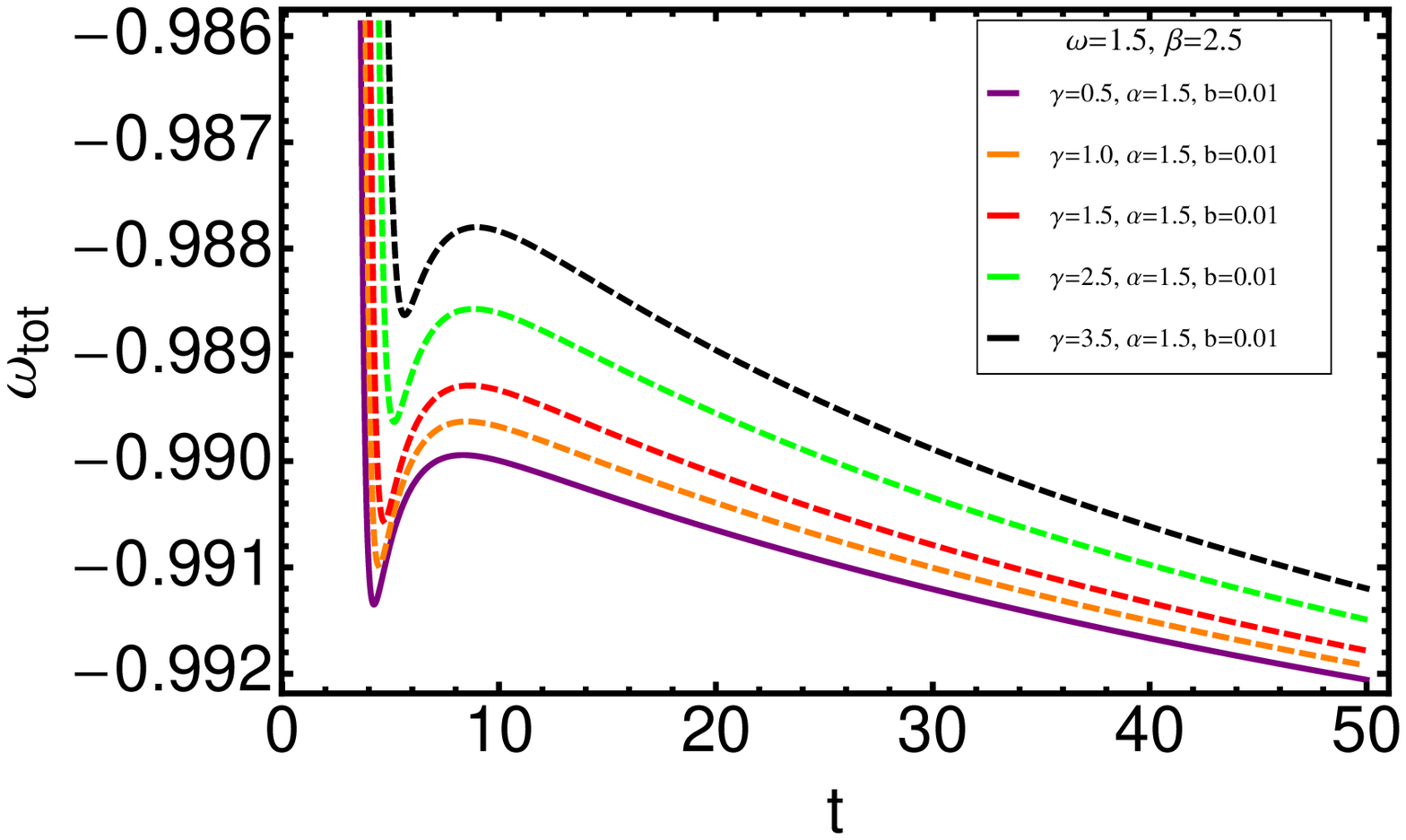}
 \end{array}$
 \end{center}
\caption{Behavior of EoS parameter $\omega_{tot}$ against $t$ for model 1.}
 \label{fig:2}
\end{figure}

\begin{figure}[h!]
 \begin{center}$
 \begin{array}{cccc}
\includegraphics[width=70 mm]{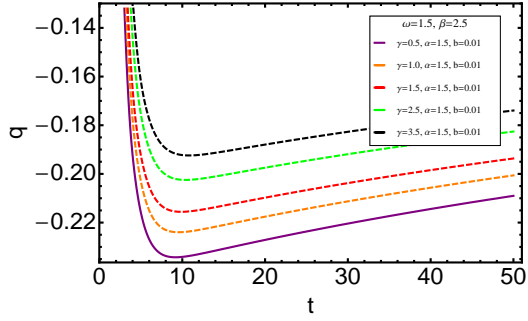}
 \end{array}$
 \end{center}
\caption{Behavior of deceleration parameter $q$ against $t$ for model 1.}
 \label{fig:3}
\end{figure}

Fig. 2 shows that total EoS yields to -1 at the late time. This is coincide with $\Lambda$CDM model. Also, Fig.3 shows that deceleration parameter in negative, while acceleration to decelerating transition happen at the initial time.\\
Fig. 3 and Fig. 4 show that $\omega_{GCCG}$ and $\omega_{Q}$ behave as total EoS.\\
Finally Fig. 6 show evolution of scalar field which is increasing function of time.

\begin{figure}[h!]
 \begin{center}$
 \begin{array}{cccc}
\includegraphics[width=70 mm]{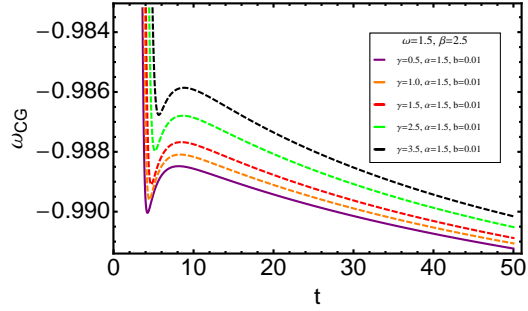}
 \end{array}$
 \end{center}
\caption{Behavior of EoS parameter of Generalized Cosmic Chaplygin Gas $\omega_{CG}$ against $t$ for model 1.}
 \label{fig:7}
\end{figure}

\begin{figure}[h!]
 \begin{center}$
 \begin{array}{cccc}
\includegraphics[width=70 mm]{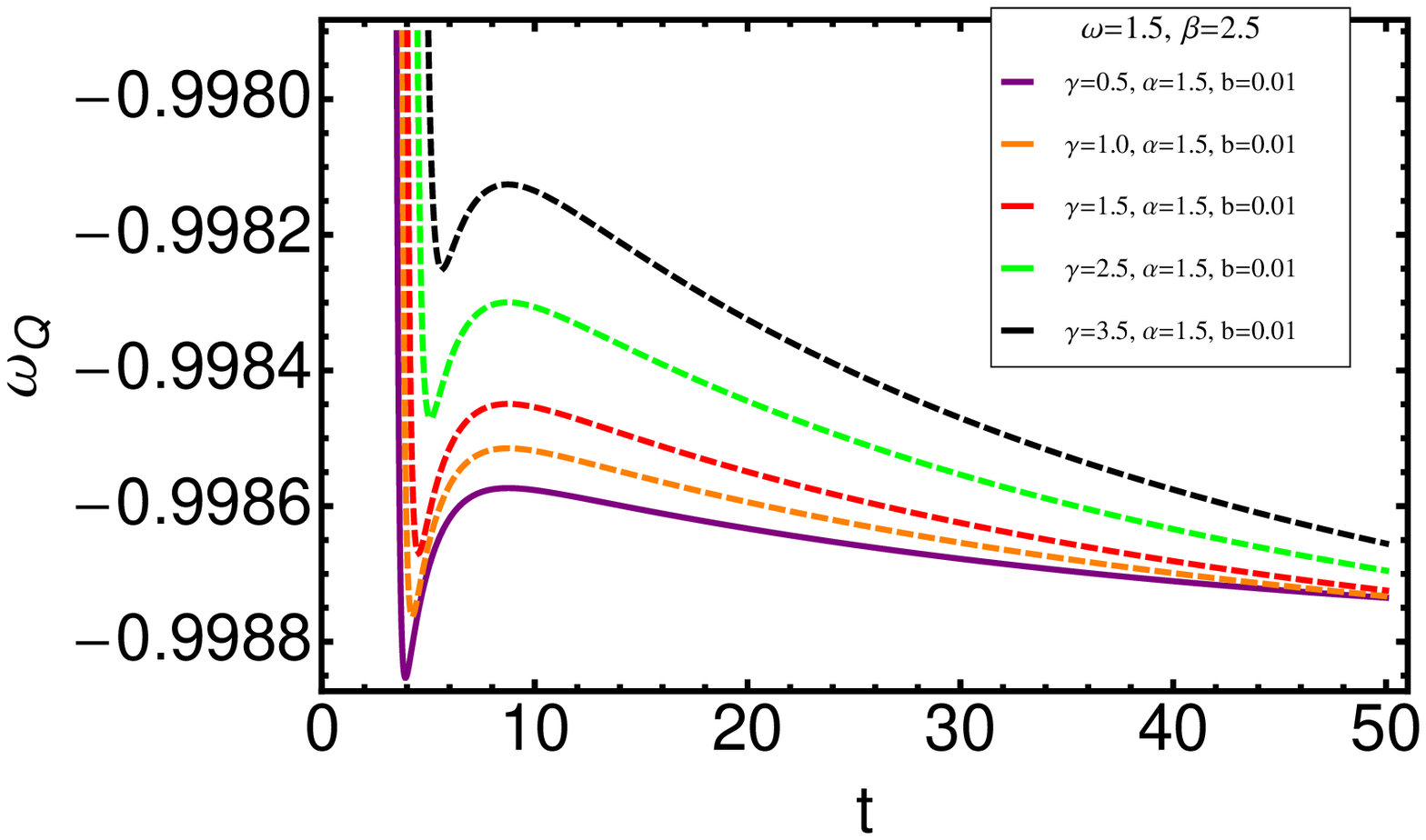}
 \end{array}$
 \end{center}
\caption{Behavior of $\omega_{Q}$ against $t$ for model 1.}
 \label{fig:8}
\end{figure}

\begin{figure}[h!]
 \begin{center}$
 \begin{array}{cccc}
\includegraphics[width=70 mm]{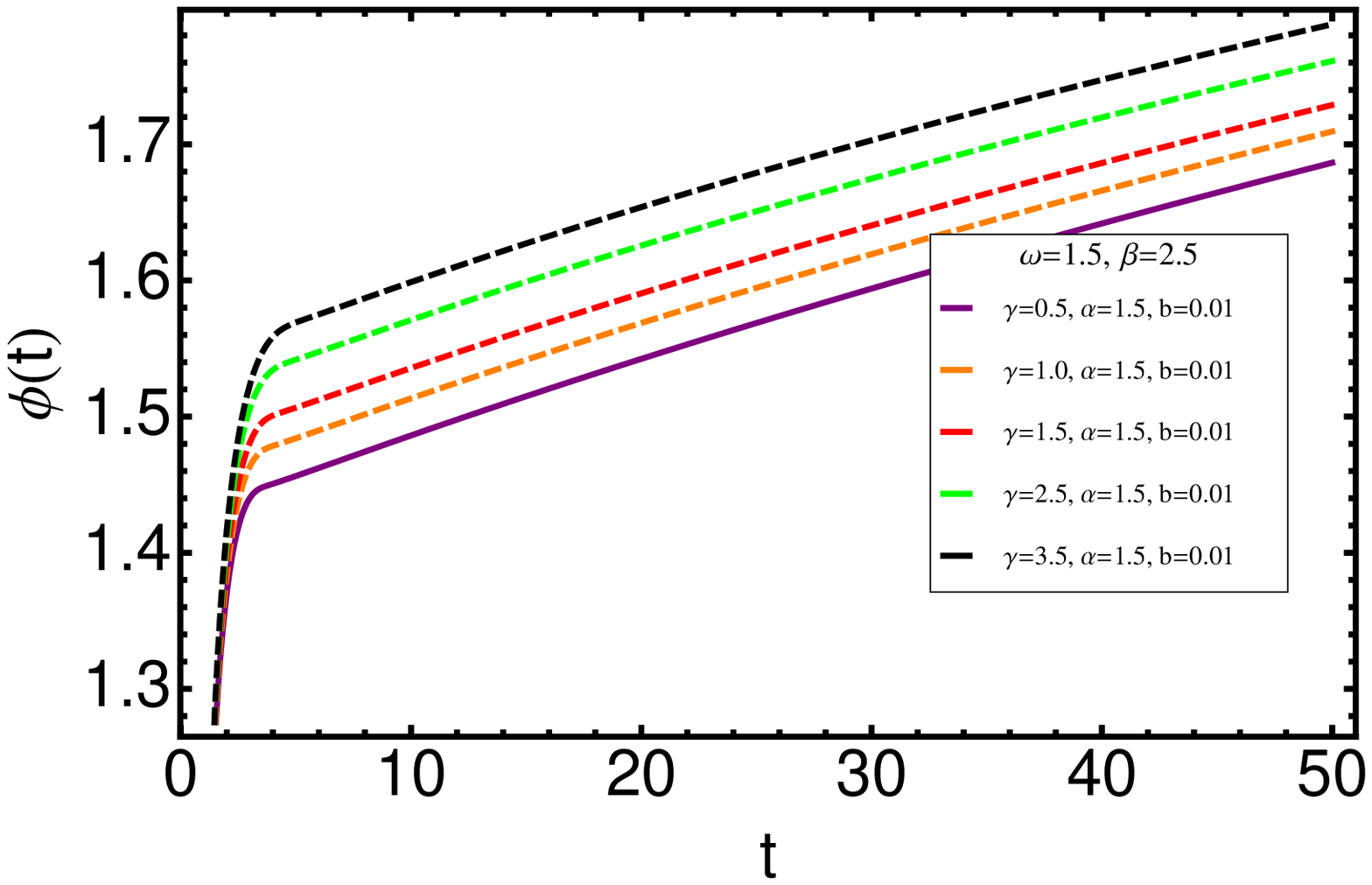}
 \end{array}$
 \end{center}
\caption{Behavior of $\phi$ against $t$ for model 1}
 \label{fig:9}
\end{figure}

\subsection{\large{Model 2}}
For this model we assumed that $V(\phi)$ is given by the equation (6). Then our numerical analysis obtain behavior of cosmological constant with time. Just as previous case we can see that Hubble expansion parameter is decreasing function of time.\\
Fig. 8 shows behavior of total EoS which yields to -1 in agreement with $\Lambda$CDM model.\\ Fig. 9 gives behavior of deceleration parameter which yields to a constant at the late time in agreement with current behavior of this parameter.\\
Figs. 10 and 11 represent $\omega_{GCCG}$ and $\omega_{Q}$ which are expected behavior. Finally, Fig. 12 shows evolution of scalar field which is increasing function of time and yields to a constant at the late time.

\begin{figure}[h!]
 \begin{center}$
 \begin{array}{cccc}
\includegraphics[width=70 mm]{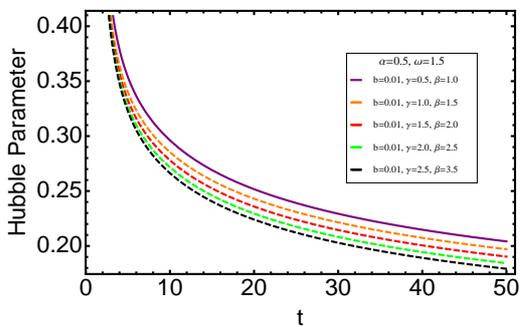}
 \end{array}$
 \end{center}
\caption{Behavior of Hubble parameter $H$ against $t$ for model 2.}
 \label{fig:4}
\end{figure}

\begin{figure}[h!]
 \begin{center}$
 \begin{array}{cccc}
\includegraphics[width=70 mm]{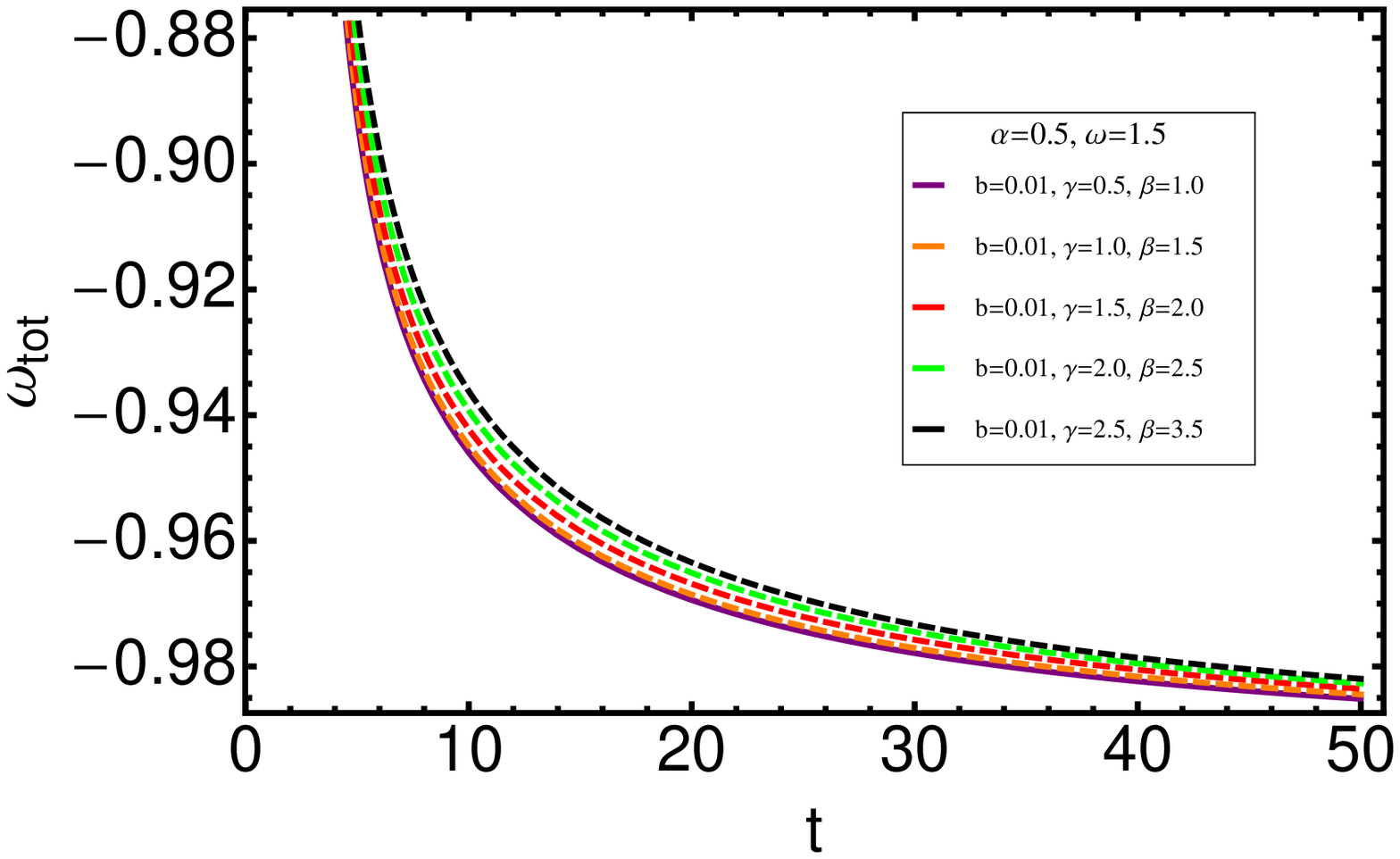}
 \end{array}$
 \end{center}
\caption{Behavior of EoS parameter $\omega_{tot}$ against $t$ for model 2.}
 \label{fig:5}
\end{figure}

\begin{figure}[h!]
 \begin{center}$
 \begin{array}{cccc}
\includegraphics[width=70 mm]{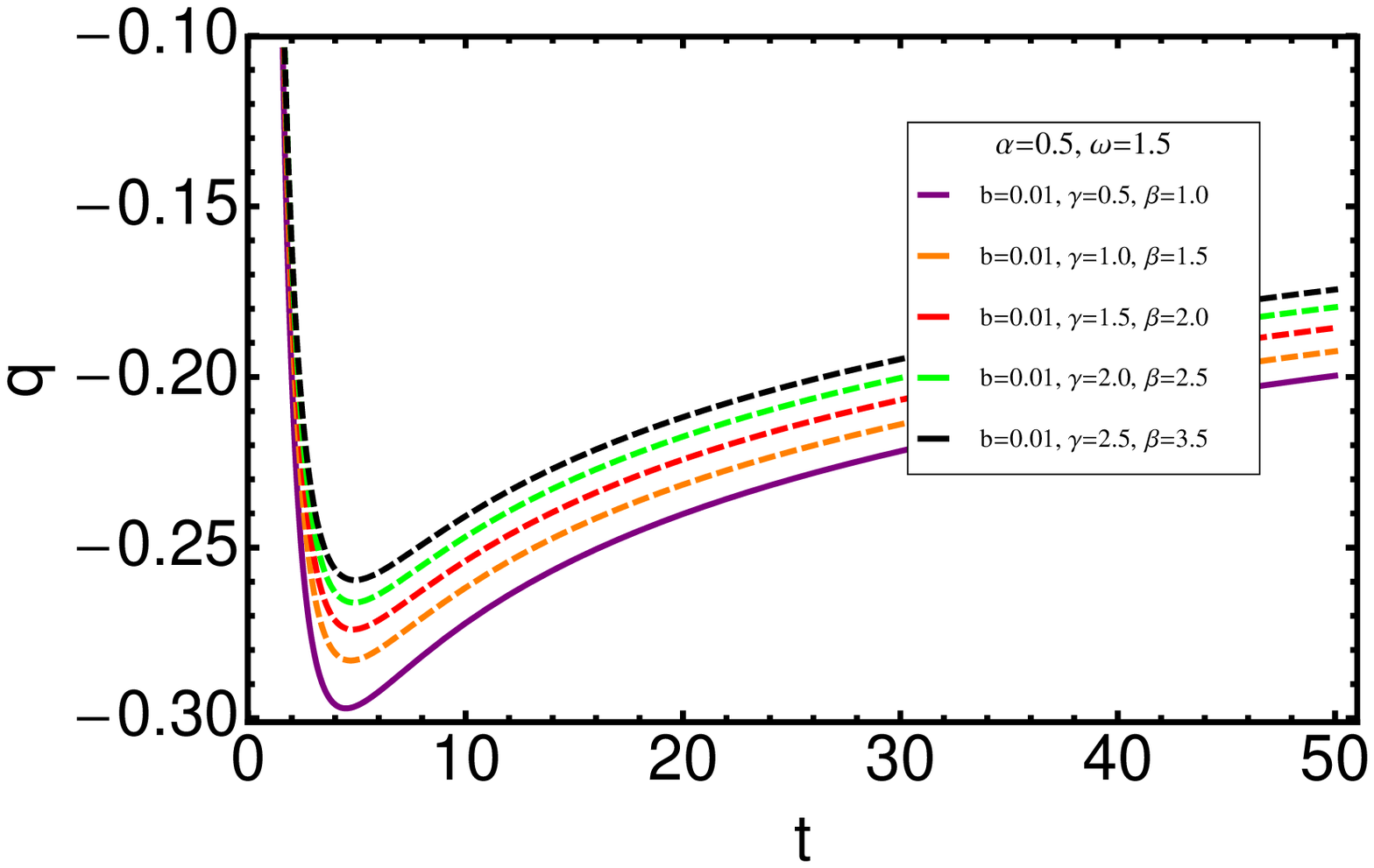}
 \end{array}$
 \end{center}
\caption{Behavior of deceleration parameter $q$ against $t$ for model 2.}
 \label{fig:6}
\end{figure}

\begin{figure}[h!]
 \begin{center}$
 \begin{array}{cccc}
\includegraphics[width=70 mm]{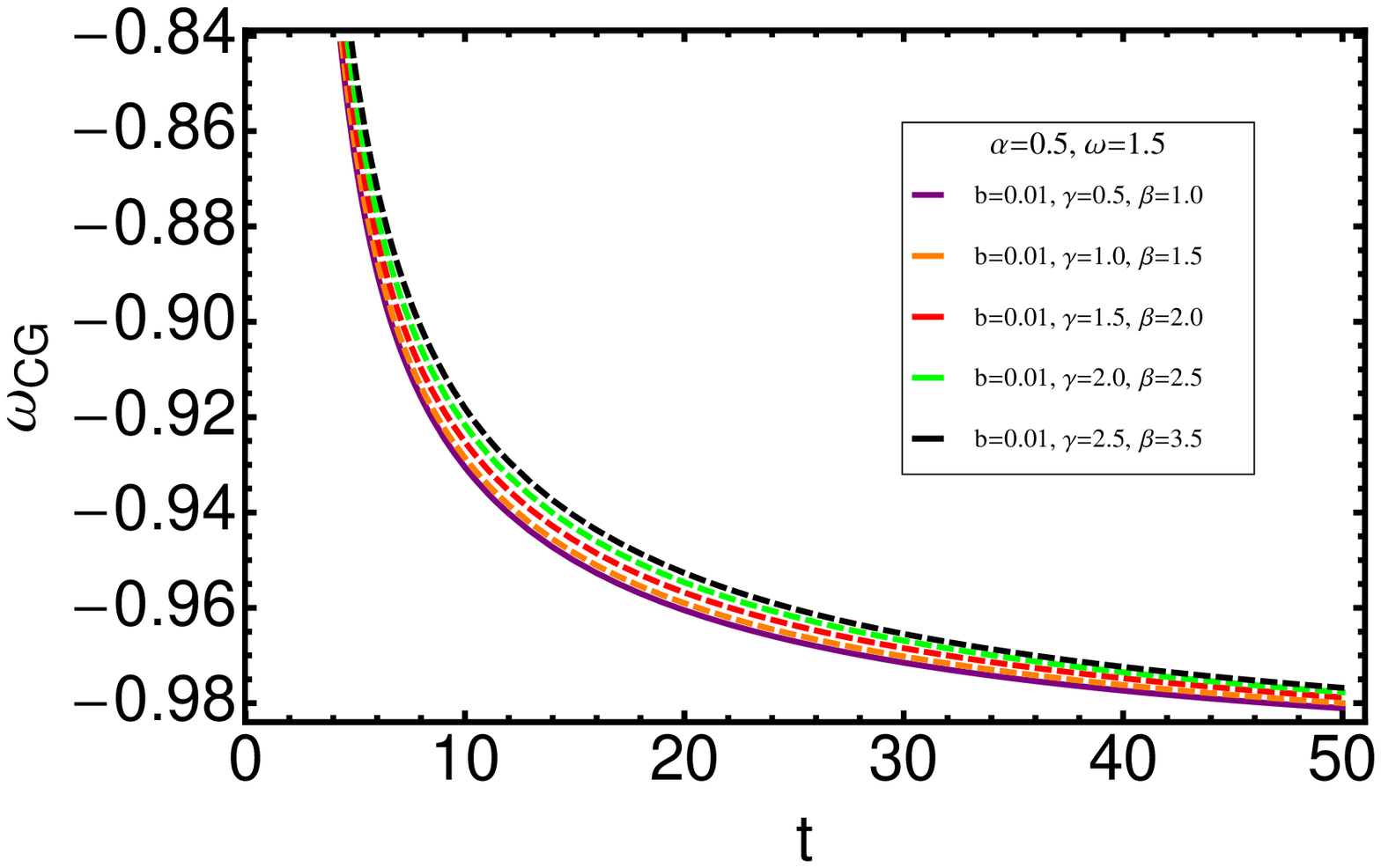}
 \end{array}$
 \end{center}
\caption{Behavior of EoS parameter of Generalized Cosmic Chaplygin Gas $\omega_{CG}$ against $t$ for model 2.}
 \label{fig:10}
\end{figure}

\begin{figure}[h!]
 \begin{center}$
 \begin{array}{cccc}
\includegraphics[width=70 mm]{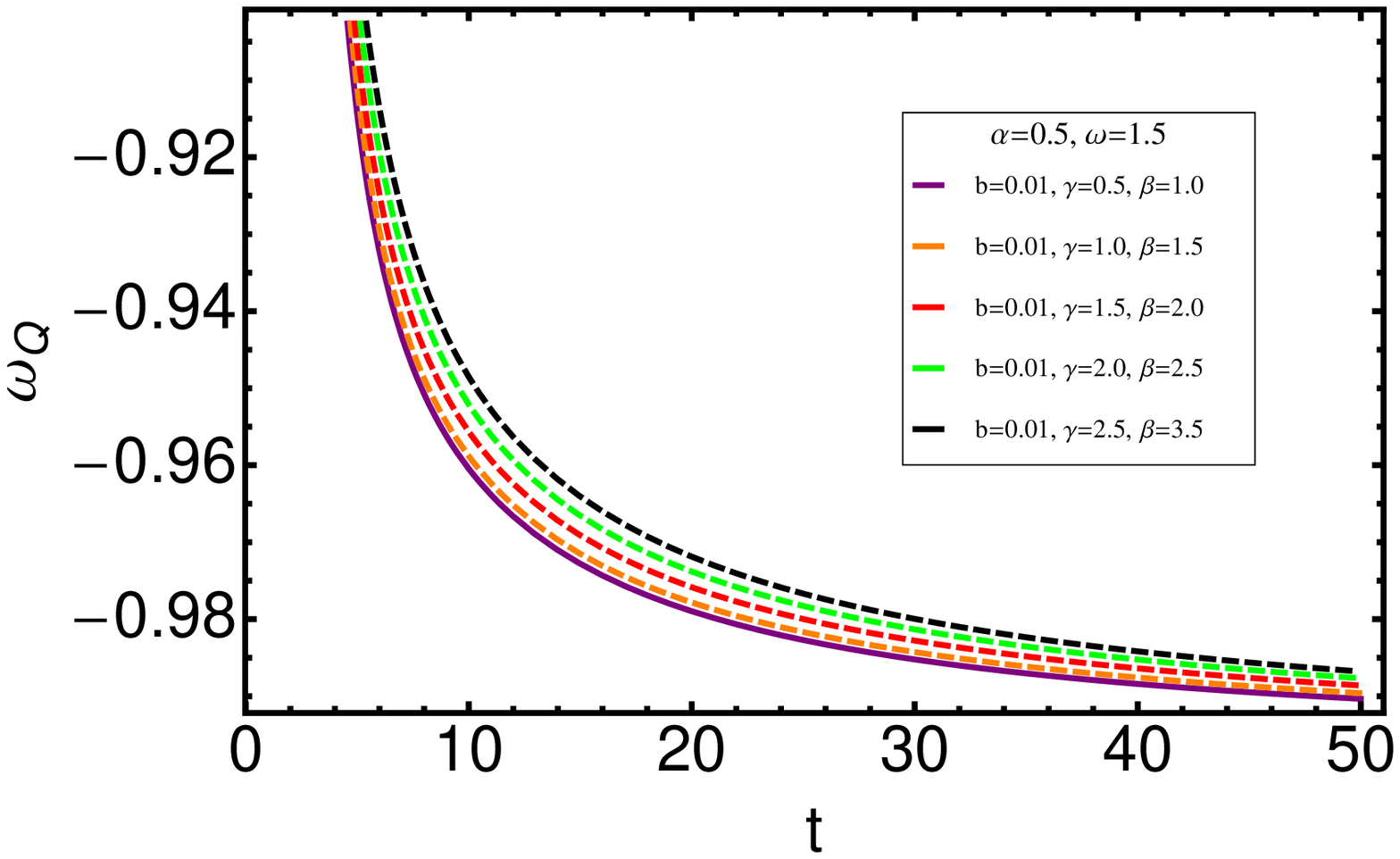}
 \end{array}$
 \end{center}
\caption{Behavior of $\omega_{Q}$ against $t$ for model 2.}
 \label{fig:11}
\end{figure}

\begin{figure}[h!]
 \begin{center}$
 \begin{array}{cccc}
\includegraphics[width=70 mm]{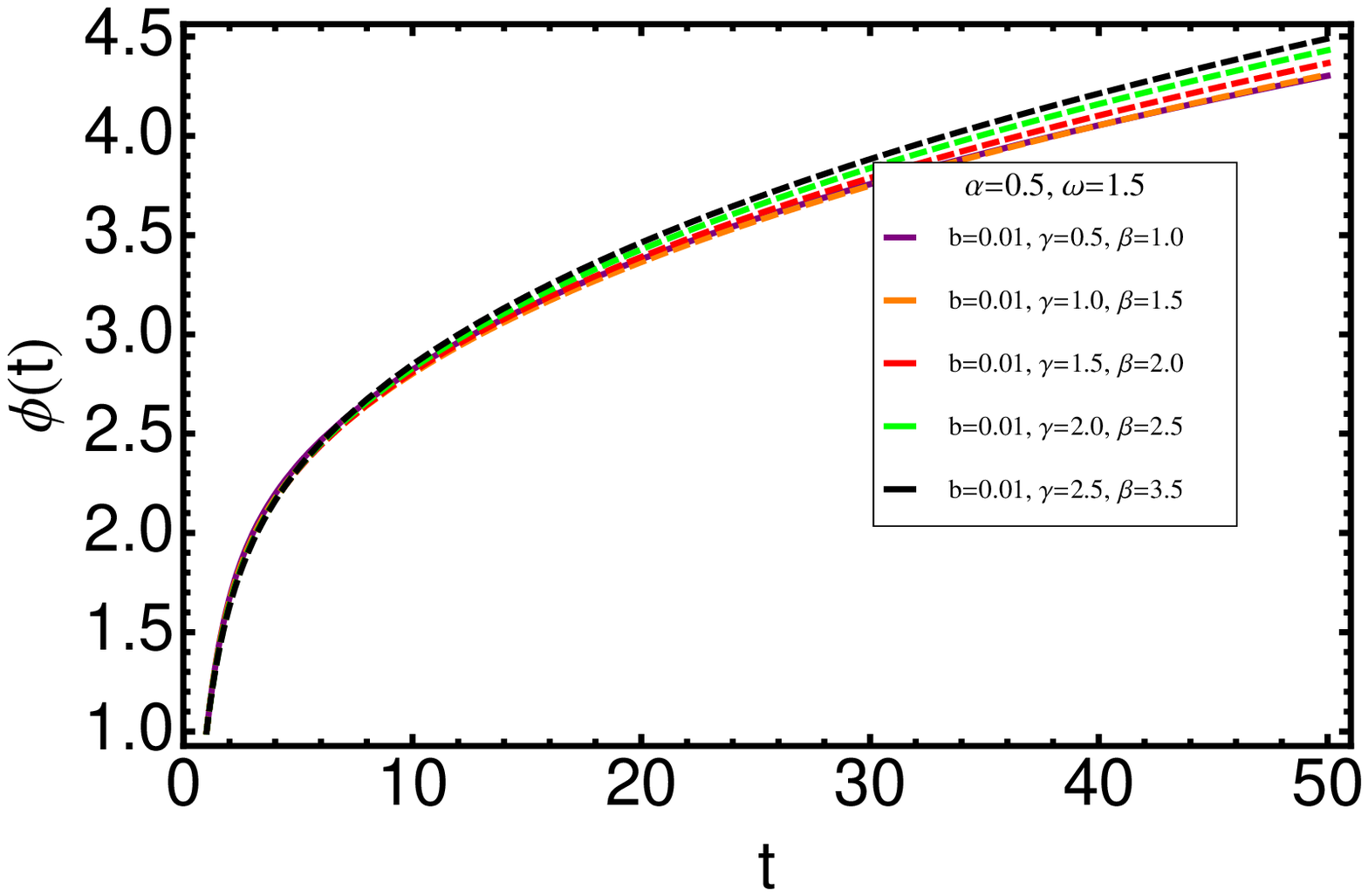}
 \end{array}$
 \end{center}
\caption{Behavior of $\phi$ against $t$ for model 2}
 \label{fig:12}
\end{figure}

\section{Conclusion}
In this paper we considered a quintessence model of dark energy which interact with generalized cosmic Chaplygin gas. Indeed we suggested the two-component Universe as a theoretical model. For the quintessence scalar field we assumed two separate possibility of potential and study both cases under interaction with generalized cosmic Chaplygin gas. An interesting assumption, which given in this paper, is consideration of variable $A$ in EoS parameter of generalized cosmic Chaplygin gas. We assumed it depend on scalar field. Our results obtained by using numerical analysis which give us behavior of some cosmological parameters such as hubble expansion parameter, deceleration parameter, EoS and scalar field. We confirmed that both cases of scalar potential have similar cosmological behavior. Both cases have decreasing Hubble expansion parameter of time, and yield to -1 EoS parameter which predicted by for example $\Lambda$CDM model. Therefore, we can propose both cases as a toy model of our Universe which agree with current observational data which tells $-1\leq\omega\leq-1/3$.
This paper may easily extend to modified cosmic Chaplygin gas [24, 25, 26].


\begin{thebibliography}{1}

\bibitem{Riess}
A.G. Riess et al. [Supernova Search Team Colloboration], Astron. J. 116 (1998) 1009
\bibitem{Perlmutter}
S. Perlmutter et al. [Supernova Cosmology Project Collaboration], Astrophys. J. 517 (1999) 565
\bibitem{Amanullah}
R. Amanullah et al., Astrophys. J. 716 (2010) 712
\bibitem{Pope}
A.C. Pope et al. Astrophys. J. 607 (2004) 655
\bibitem{5}
C. Armendariz-Picon, V. Mukhanov and P.J. Steinhardt, Phys. Rev.
Lett. 85 (2000) 4438
\bibitem{6}
A.Y. Kamenshchik, U. Moschella, and V. Pasquier, Phys. Lett. B 511
(2001) 265
\bibitem{7}
M.R. Setare, Phys. Lett. B648 (2007) 329
\bibitem{8}
M.R. Setare, Phys. Lett. B654 (2007) 1
\bibitem{9}
V. Sahni and A.A. Starobinsky, Int. J. Mod. Phys. D 9 (2000) 373
\bibitem{10}
S. Nobbenhuis, Found. Phys. 36 (2006) 613
\bibitem{P11}
J. Sadeghi, B. Pourhassan and Z.A. Moghaddam, Int. J. Theor. Phys. 53 (2014) 125
\bibitem{P12}
N. Bilic, G.B. Tupper, and R.D. Viollier, Phys. Lett. B 535 (2002) 17
\bibitem{P13}
D. Bazeia, Phys. Rev. D 59 (1999) 085007
\bibitem{P14}
L. Xu, J. Lu, Y. Wang, Eur. Phys. J. C72 (2012) 1883
\bibitem{P15}
H. Saadat and  B. Pourhassan, "Effect of Varying Bulk Viscosity on Generalized Chaplygin Gas", IJTP DOI: 10.1007/s10773-013-1913-8
\bibitem{P16}
Xiang-Hua Zhai et al. Int. J. Mod. Phys. D15 (2006) 1151
\bibitem{P17}
Y. D. Xu et al. Astrophys Space Sci 337 (2012) 493
\bibitem{P18}
H. Saadat and B. Pourhassan, Astrophys Space Sci. 343 (2013) 783
\bibitem{P19}
H. Saadat and H. Farahani,  Int. J. Theor. Phys. 52 (2013) 1160 DOI 10.1007/s10773-012-1431-0
\bibitem{P20}
A. R. Amani and B. Pourhassan, Int. J. Theor. Phys. 52 (2013) 1309
\bibitem{P21}
U. Debnath, A. Banerjee, and S. Chakraborty, Class. Quantum Grav.
21 (2004) 5609
\bibitem{P22}	
J. Naji, B. Pourhassan, A.R. Amani, International Journal of Modern Physics D Vol. 23, No. 1 (2013) 1450020
\bibitem{P23}
H. Saadat and  B. Pourhassan, Astrophysics and Space Science 344 (2013) 237
\bibitem{24}
J. Sadeghi and H. Farahani, Astrophysics and Space Science 347 (2013) 209
\bibitem{25}
B. Pourhassan, International Journal of Modern Physics D Vol. 22, No. 9 (2013) 1350061
\bibitem{26}
J. Sadeghi, M. Khurshudyan, B. Pourhassan, H. Farahani, IJTP DOI: 10.1007/s10773-013-1881-z
\end{thebibliography}
\end{document}